\documentclass[10pt]{iopart}

\usepackage{graphicx}
\usepackage{hyperref}
\usepackage{xcolor}
\usepackage{multicol}
\usepackage{upgreek}
\expandafter\let\csname equation*\endcsname\relax
\expandafter\let\csname endequation*\endcsname\relax
\usepackage{amsmath}
\usepackage{MnSymbol}
\usepackage[noadjust]{cite}

\begin{document}

\title[Indirect measurement of atomic magneto-optical rotation via Hilbert Transform]{Indirect measurement of atomic magneto-optical rotation via Hilbert transform}

\author{Jack~D~Briscoe,  Danielle~Pizzey, Steven~A~Wrathmall and Ifan~G~Hughes}

\address{\textit{Department of Physics, Durham University, South Road, Durham, DH1 3LE, United Kingdom.}}

\ead{jack.d.briscoe@durham.ac.uk}

\begin{abstract} 
The Kramers-Kronig relations are a pivotal foundation of linear optics and atomic physics, embedding a physical connection between the real and imaginary components of any causal response function. A mathematically equivalent, but simpler, approach instead utilises the Hilbert transform. In a previous study, the Hilbert transform was applied to absorption spectra in order to infer the sole refractive index of an atomic medium in the absence of an external magnetic field. The presence of a magnetic field causes the medium to become birefringent and dichroic, and therefore it is instead characterised by two refractive indices. In this study, we apply the same Hilbert transform technique to independently measure both refractive indices of a birefringent atomic medium, leading to an indirect measurement of atomic magneto-optical rotation. Key to this measurement is the insight that inputting specific light polarisations into an atomic medium induces absorption associated with only one of the refractive indices. We show this is true in two configurations, commonly referred to in literature as the Faraday and Voigt geometries, which differ by the magnetic field orientation with respect to the light wavevector. For both cases, we measure the two refractive indices independently for a Rb thermal vapour in a $0.6~\mathrm{T}$ magnetic field, finding excellent agreement with theory. This study further emphasises the application of the Hilbert transform to the field of quantum and atomic optics in the linear regime.

\end{abstract}


\ioptwocol

\section{Introduction}
The Kramers-Kronig (KK) relations are used extensively across a wide range of disciplines of physics~\cite{lovell1974application, roessler1965kramers, bakry2018using, horsley2015spatial, sai2020designing, darwish2019deposition, o1981kramers, szabo2010unique, macdonald1985application, tanner2015use}, including signal reconstruction in the optical~\cite{mecozzi2016kramers} and terahertz regimes~\cite{harter2020generalized}, measuring the scattering properties of blood~\cite{faber2004oxygen}, and  holographic imaging~\cite{baek2021intensity, huang2022high}.  The reason for their ubiquity is made clear by their use. For any complex function which is linear, causal and analytic in the upper half-plane~\cite{toll1956causality}, the KK relations map the real component of that function to its imaginary counterpart~\cite{kronig1926theory, kramers1927diffusion}, and therefore knowledge of one component means inferring the other indirectly. The KK relations are a pivotal foundation of linear optics and atomic physics, where one is interested in the behaviour of light as it traverses through a dispersive medium~\cite{demtroder1982laser}. The underlying physics is governed by the complex refractive index~\cite{adams2018optics}; the real component is responsible for the dispersive properties of the medium, i.e. light slowing~\cite{camacho2007wide}, while the imaginary component describes absorption of light~\cite{siddons2008absolute}. Absorption spectroscopy is simple to perform, and consequently, both fundamental and applied studies are exhaustive~\cite{pizzey2022laser}. On the other hand, light dispersion, i.e. the frequency dependence of the refractive index~\cite{purves2004refractive}, is difficult to measure, often requiring complex interferometric setups~\cite{xiao1995measurement, keaveney2012maximal, stavenga2013quantifying} that are susceptible to systematic errors such as mechanical vibrations and thermal drifts. The KK relations bypass this method by offering a simple way to indirectly measure dispersion via less complex absorption spectroscopy~\cite{whittaker2015hilbert}. 

The main difficulty with the KK relations is that they are computationally complex. From an analysis perspective, one must take care in avoiding the poles of the integrand (see equation~\ref{eqn: generalKK}). The calculation is also slow, since only a single value of the real refractive index is extracted when integrating over a large frequency range of its imaginary counterpart. This motivates application of the Hilbert transform (HT), which is a mathematical analogue of the KK relations~\cite{king_2009}. In fact, the KK relations form a HT-pair, and can both be derived in a similar manner via complex contour integration~\cite{boyd2020}. There are many applications of the HT~\cite{benitez2001use, volkov2020situ, li2016phase, zhu2012electron, davis2000image, mishnev1993discrete}, particularly in signal processing~\cite{feldman2011hilbert}. Consequently, the HT is a staple of many signal processing software packages, such as Python's \verb|SciPy|~\cite{virtanen2020scipy}. These typically utilise fast Fourier transforms; not only does this mean the calculation avoids dealing with poles, but it is also makes the transformation much faster than the alternative KK relations. The HT is therefore the preferred transformation method\footnote{For a treatise on HT applications, we refer the reader to Chapter 23 in volume II of~\cite{king_2009}; for KK relations in the context of both linear and non-linear optics research, see~\cite{lucarini2005kramers}.}.

A previous study utilised the HT on an atomic system in the absence of a magnetic field to indirectly measure its sole refractive index~\cite{whittaker2015hilbert}. Magnetic fields are used extensively by atomic physics researchers, with applications found in both fundamental~\cite{breit1931measurement, tremblay1990absorption, umfer1992investigations, windholz1985zeeman, windholz1988zeeman, ponciano2020absorption, pizzey2021tunable, alqarni2023device, trenec2011permanent} and applied~\cite{sutter2020recording, staerkind2023precision, staerkind2023high, higgins2021electromagnetically, olsen2011optical, ciampini2017optical, auzinsh2022wide, sargsyan2012hyperfine} studies. Of interest in this study is the property that an atomic medium becomes birefringent when subject to an external magnetic field, and therefore the system exhibits two refractive indices $n_{1,2}$~\cite{adams2018optics}. This property is exploited by atomic devices such as optical isolators~\cite{weller2012optical, aplet1964faraday} and atomic filters~\cite{dick1991ultrahigh, yeh1982dispersive, gerhardt2018anomalous, kiefer2014faraday, logue2022better}, where differential refraction i.e. birefringence, causes a plane of polarisation rotation in the vicinity of atomic resonances~\cite{uhland2023build}. Consequently, these devices are excellent at rejecting undesirable light~\cite{yin2022using}. 

In this study, we show that the birefringence an atomic medium exhibits is a function of its atom-light geometry (i.e. the relative angle between the light wavevector $\vec{k}$ and magnetic vector $\vec{B}$), and solve for two cases of analytic magneto-optical rotation. These are the Faraday geometry~\cite{faraday1846experimental, zentile2014hyperfine, chang2017faraday, aplet1964faraday, gerhardt2018anomalous}, where $\vec{k}$ is parallel to $\vec{B}$ ($\vec{k} \parallel \vec{B}$), and the Voigt geometry~\cite{voigt1899theorie, ponciano2020absorption, briscoe2023voigt, kudenov2020dual, liu2023atomic, muroo1994resonant, mottola2023electromagnetically, mottola2023quantum}, where $\vec{k}$ is perpendicular to $\vec{B}$ ($\vec{k} \perp \vec{B}$). The HT technique is applied to both of these systems, leading to an indirect measurement of the two refractive indices $n_{1,2}$ for each geometry via simple absorption spectroscopy. Notably, while previous research has focused on measuring total magneto-optical rotation~\cite{budker2002resonant, auzinsh2010optically, budker2000sensitive, carr2020measuring, maxwell2021white, wu1986optical, wolfenden1990use, edwards1995precise, edwards1995magneto, weller2012measuring, zentile2014hyperfine, kemp2011analytical, siddons2010optical, siddons2009gigahertz}, to our knowledge, the two individual refractive indices have never been separately measured in studies of atomic thermal vapours\footnote{Reference~\cite{zentile2014hyperfine} discusses a method to measure both refractive indices of an atomic medium in the Faraday geometry, but this requires a large magnetic field, small optical depths, and is only valid in the vicinity of atomic resonances.}. Measuring the difference $\Delta n = n_{2} - n_{1}$ ultimately provides an indirect measurement of atomic magneto-optical rotation via the HT.

To extract $n_{1,2}$ independently, one needs to understand how light is propagated through an atomic medium subject to an external magnetic field. This involves solving a dispersion equation for both the refractive indices and propagation eigenmodes of any atom-light geometry~\cite{rotondaro2015generalized, palik1970infrared, keaveney2018elecsus}. We show that by inputting light into an atomic medium that matches one of its propagation eigenmodes, the light exhibits an atom-light interaction that is only associated with one of the refractive indices. This assumes the medium has orthogonal eigenmodes, which the Faraday and Voigt geometries do. By measuring this interaction via absorption spectroscopy, we can infer the real component of both refractive indices via the HT. 

The paper is structured as follows: Section~\ref{sec: KK HT} shows the connection between the KK relations and the HT, mapping the imaginary component of the electric susceptibility to its real counterpart; the method to calculate refractive index from absorption spectrum is shown in Section~\ref{sec: absorptionToN}; independent measurement of the two refractive indices of a birefringent atomic medium is presented in Section~\ref{sec: experiment} for both the Faraday and Voigt geometries, followed by a conclusion in Section~\ref{sec: conclusion}.

\section{Kramers-Kronig relations and the Hilbert transform} \label{sec: KK HT}

The KK relations for the complex susceptibility $\chi$ can be derived using complex analysis~\cite{king_2009, boyd2020}. Assuming $\chi$ is analytic in the upper half plane\footnote{Since $\chi$ is a linear response function, it is also causal. Titchmarsh shows causality and analyticity in the upper half plane are one and the same~\cite{titchmarsh1948introduction}; also, see Toll~\cite{toll1956causality}.}, careful choice of contour and application of both Cauchy's residue and integral theorems leads to

\begin{equation}
\chi (\omega) =  \frac{1}{\mathrm{i}\pi}\mathcal{P}\int_{-\infty}^{\infty}\frac{\chi (\omega ')}{\omega ' - \omega}\rm{d}\omega ' \,,
\label{eqn: generalKK}
\end{equation}

\noindent where $\mathcal{P}\int$ represents a Cauchy principal value integral~\cite{kanwal2013linear} and $\omega$ is the angular frequency of an incident electromagnetic field. The HT is one of many integral transforms that takes the form $g(x) = \int^{b}_{a} k(x,y) f(y) dy$, where $k(x,y)$ is known as the kernel~\cite{kanwal2013linear, king_2009}. For any function $f(x)$, application of the HT is defined by~\cite{king_2009}

\begin{equation}
\mathcal{H}[f(x)] =  \frac{1}{\pi}\mathcal{P}\int_{-\infty}^{\infty}\frac{f(y)}{x - y}\rm{d}y \,,
\label{eqn: generalHT}
\end{equation}

\noindent which has a structure analogous to equation~\ref{eqn: generalKK}. As is shown in~\cite{whittaker2015hilbert, king_2009}, it is therefore simple to infer $\chi^{\mathcal{R}}$ from $\chi^{\mathcal{I}}$ using the HT~\cite{adams2018optics}

\begin{equation}
\chi^{\mathcal{R}}(\omega) = -\mathcal{H}[\chi^{\mathcal{I}} (\omega)] \,,
\label{eqn: chiHT}
\end{equation}

\noindent where the minus sign is a consequence of the definitions used in equations \ref{eqn: generalKK} and \ref{eqn: generalHT}. We note that superscript notation is used throughout this paper to denote the real ($\mathcal{R}$) and imaginary ($\mathcal{I}$) components of complex variables. In practice, the HT method (equation~\ref{eqn: chiHT}) can be implemented in Python using \verb|SciPy|'s signal package~\cite{virtanen2020scipy}.

In general, the HT (and KK relations) apply to any dispersive medium in the linear optics regime; in our work, we are interested in the complex susceptibility of two-level atoms subject to weak probe monochromatic radiation~\cite{zentile2015elecsus}. For atomic systems subject to an external magnetic field, we no longer consider $\chi$ as the total susceptibility of the medium, but instead separate $\chi$ into three components based on atomic transitions. These components are labelled $\chi_{+1}$, $\chi_{-1}$ and $\chi_{0}$; the subscripts are chosen to match the change in projection quantum number $\Delta m_{l}$ associated with electric-dipole selection rules~\cite{foot2004atomic}. We start by calculating the atomic transition frequencies using a matrix representation; for more details, see~\cite{wellerThesis, zentile2015elecsus}. The matrix representation allows our theoretical model to filter transitions by $\Delta m_{l}$, such that the frequency dependent lineshape~\cite{siddons2009off} of each individual transition can be summed together and assigned to the correct component of $\chi$. We illustrate the components $\chi_{q}$ in figure~\ref{fig: HilbertMethod} (bottom row), where $q = \Delta m_{l} \in \{+1, -1, 0\}$ represents each electric-dipole selection rule, and $\chi^{\mathcal{I}}$ exhibits a Lorentzian lineshape. For an atomic thermal vapour, the lineshape of $\chi^{\mathcal{I}}$ is the Voigt profile~\cite{siddons2008absolute, pizzey2022laser}, which maps to a Doppler-broadened dispersion curve~\cite{zentile2014hyperfine} via the KK relations. This gives the lineshape of $\chi^{\mathcal{R}}$, which for a two-level atom in the weak probe regime is also found using theory~\cite{siddons2008absolute}. For details of how the components $\chi_{q}$ are integral to light propagation, see \ref{App1}.

We postulate that similar relationships to equation~\ref{eqn: chiHT} hold true for all components of $\chi$

\begin{equation}
\chi^{\mathcal{R}}_{q}(\omega) = -\mathcal{H}[\chi^{\mathcal{I}}_{q} (\omega)] \,.
\label{eqn: chiHTcomponents}
\end{equation}

\noindent The significance of equation~\ref{eqn: chiHTcomponents} is demonstrated in the next section.

\section{From absorption spectrum to refractive index} \label{sec: absorptionToN}

\begin{figure*}[t]
\centering
{\includegraphics[width=0.975\linewidth]{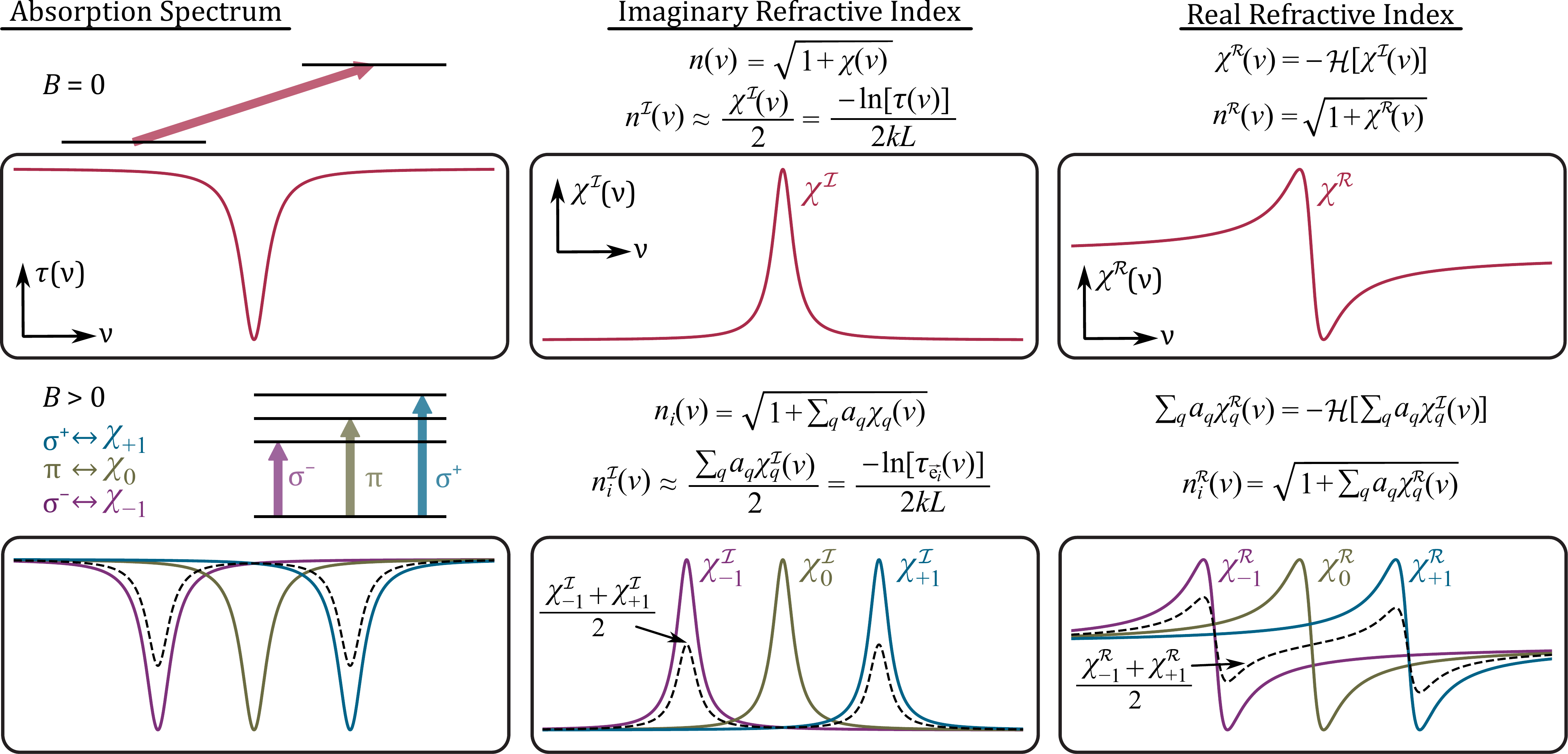}}
\caption{Figure showing the process to transform from experimental absorption spectrum into the real refractive indices of a two-level atomic medium, assuming the linear optics regime. In the zero field case (top row), the imaginary electric susceptibility $\chi^{\mathcal{I}}$ is calculated via simple rearrangement of the Beer-Lambert transmission law $\tau(\nu) = \mathrm{exp}(-\chi^{\mathcal{I}}k L)$, where $k$ is the wavevector, $L$ is the medium length, and $\nu$ is linear detuning. A Hilbert transform maps $\chi^{\mathcal{I}} \leftrightarrow \chi^{\mathcal{R}}$, which is then used to calculate the real refractive index $n^{\mathcal{R}}$. In the presence of an external magnetic field (bottom row), the excited state energy level splits via the Zeeman effect. Consequently, the lineshape is described by three components $\chi_{q}$, where $q = \Delta m_{l} \in \{-1, 0, +1\}$ corresponds to electric-dipole transition rules. The general refractive index equations are shown, and are linear combinations of $\chi_{q}$ weighted by real constants $a_{q}$. The atomic medium is now birefringent; each refractive index $n_{i}$ is isolated via inputting a light polarisation into the atomic medium which is one of its propagation eigenmodes $\vec{e}_{i}$, assuming the eigenmodes of the system are orthogonal (see text). Consequently, the output transmission $\tau_{\vec{e}_{i}}$ is only a function of the associated refractive index $n_{i}$. Energy level schematics represent toy atomic models.}
\label{fig: HilbertMethod}
\end{figure*}

Figure~\ref{fig: HilbertMethod} shows the general principle to measure the real refractive index from a simple absorption spectrum. The top row shows the zero $B$-field case, as described in~\cite{whittaker2015hilbert}. An absorption spectrum measures the transmission $\tau$ of the laser beam after it traverses an atomic medium, and in general, $\tau = I(L)/I_{0}$, where $I_{0}$ is the incident intensity, and $I(L)$ is the intensity after a propagation distance $L$. In the absence of a magnetic field, $\tau$ is calculated using the Beer-Lambert law; from literature, we find $\tau(\nu) = \mathrm{exp}(-\chi^{\mathcal{I}}k L)$~\cite{pizzey2022laser}. This equation can then be rearranged to infer the imaginary component of the electric susceptibility

\begin{equation}\label{eqn: electricSusceptibilityIntensity}
\chi^{\mathcal{I}}(\nu) = \frac{-\mathrm{ln}[\tau(\nu)]}{k L} \,,
\end{equation}
 
\noindent where $k = \lvert \vec{k} \rvert$, and $\tau$ is measured as a function of linear detuning $\nu$ from the weighted line centre of the absorption spectrum~\cite{siddons2008absolute}. Application of equation~\ref{eqn: chiHT} maps $\chi^{\mathcal{I}}$ to its real counterpart, where it is used to infer the refractive index $n^{\mathcal{R}} = \sqrt{1 + \chi^{\mathcal{R}}}$ \cite{adams2018optics}. This provides the simplest picture of how one can measure light dispersion from light absorption.

We extend the work presented in~\cite{whittaker2015hilbert} to two analytic cases where atoms in a thermal vapour are subject to an external magnetic field. These cases are referred to as the Faraday and Voigt geometries, where the relative angle between $\vec{k}$ and $\vec{B}$ is $\theta_{B} = 0^{\circ}$ and $\theta_{B} = 90^{\circ}$ respectively. We model the behaviour of light as it traverses an atomic medium in any geometry using a modified version of the open-source computer program $ElecSus$~\cite{zentile2015elecsus, keaveney2018elecsus}. \ref{App1} covers the method $ElecSus$ uses to propagate an input electric field through an atomic medium; for further details, we refer the reader to~~\cite{rotondaro2015generalized, keaveney2018elecsus, logue2023exploiting} and references therein. In short, we solve a dispersion equation to determine the refractive indices $n_{1,2}$ and corresponding propagation eigenmodes $\vec{e}_{1, 2}$ of the birefringent and dichroic atomic medium. The eigenmodes are used to rotate the incident electric field $\vec{\mathcal{E}}_{\mathrm{in}}$ into the natural coordinate system of the atom-light geometry, and the refractive indices are used to propagate the light through the atomic medium. The output electric field $\vec{\mathcal{E}}_{\mathrm{out}}$ is calculated after an inverse rotation back into Cartesian coordinates, once again using the eigenmodes. From ~\ref{App1}

\begin{subequations} \label{eqn: FaradayPropagationEigenmodes}
\begin{align}
& \left.\begin{array}{l}
    n_{1, \mathcal{F}} \approx 1 + \chi_{+1}/2 \\
    \end{array}\right\} \quad \vec{e}_{1, \mathcal{F}} = 1/\sqrt{2}\begin{pmatrix} 1 & \mathrm{i} 
    \end{pmatrix}^{\mathrm{T}} \\
& \left.\begin{array}{l}
    n_{2, \mathcal{F}} \approx 1 + \chi_{\mathrm{-1}}/2 \\
    \end{array}\right\} \quad \vec{e}_{2, \mathcal{F}} = 1/\sqrt{2}\begin{pmatrix} 1 & -\mathrm{i} 
    \end{pmatrix}^{\mathrm{T}} \,,
\end{align}
\end{subequations}

\noindent for Faraday ($\mathcal{F}$), whereas for Voigt ($\mathcal{V}$) 

\begin{subequations} \label{eqn: VoigtPropagationEigenmodes}
\begin{align}
& \left.\begin{array}{l}
    n_{1, \mathcal{V}} \approx 1 + \chi_{\mathrm{0}}/2 \\
    \end{array} \quad\quad\quad  \right\} \quad \vec{e}_{1, \mathcal{V}} = \begin{pmatrix} 1 & 0 
    \end{pmatrix}^{\mathrm{T}} \\
& \left.\begin{array}{l}
    n_{2, \mathcal{V}} \approx 1 + (\chi_{\mathrm{+1}} + \chi_{\mathrm{-1}})/4 \\
    \end{array}\right\} \quad \vec{e}_{2, \mathcal{V}} = \begin{pmatrix} 0 &  1 
    \end{pmatrix}^{\mathrm{T}} \,.
\end{align}
\end{subequations}

\noindent The eigenmodes indicate the type of birefringence the medium will exhibit; for Faraday, light exhibits circular birefringence, whereas for Voigt, light exhibits linear birefringence. The refractive indices indicate the types of transitions an atomic medium will exhibit, since they are functions of the different components of the electric susceptibility $\chi_{q}$. In the Faraday geometry, $
\sigma^{\pm}$ transitions ($\Delta m_{l} = \pm 1$) are induced by left- and right-hand circularly polarised light (equation~\ref{eqn: FaradayPropagationEigenmodes}), and therefore an atomic medium orientated with $\theta_{B} = 0$ will not exhibit $\pi$ transitions ($\Delta m_{l} = 0$)\footnote{This is only true when light is a transverse wave~\cite{adams2018optics}.}. In the Voigt geometry, $\pi$ transitions are induced by horizontal linear light, whereas $\sigma^{\pm}$ transitions are induced simultaneously by vertical linear light (equation~\ref{eqn: VoigtPropagationEigenmodes}). This specific example is illustrated in figure~\ref{fig: HilbertMethod} (bottom row); the two components $\chi_{+1}^{\mathcal{I}}$ and $\chi_{-1}^{\mathcal{I}}$ are summed and weighted by factors of a half. Since the HT is linear, the total lineshape can then be Hilbert transformed to infer the sum of their real components, and therefore $n_{2, \mathcal{V}}$.

We can isolate an interaction that only depends on one of the two refractive indices by inputting light into an atomic medium which is one of its propagation eigenmodes. This statement is proved in \ref{App2}, and assumes orthogonal eigenmodes\footnote{We learn from~\cite{logue2023exploiting} that in general an atomic medium exhibits non-orthogonal eigenmodes, and that Faraday ($\theta_{B} = 0^{\circ}$) and Voigt ($\theta_{B} = 90^{\circ}$) are the only exceptions.}. From the same appendix, the output electric field after the medium is given by

\begin{equation}\label{eqn: 2D propagationInputEigenmode1}
\vec{\mathcal{E}}_{\mathrm{out}} = f(n_{i})  \begin{pmatrix}
e_{i1} \\
e_{i2} \\
\end{pmatrix} = f(n_{i})\mathcal{E}_{0}\vec{e}_{i} \,,
\end{equation}

\noindent where $e_{ij}$ are components of the normalised eigenmode $\vec{e}_{i} = (e_{i1}, e_{i2})^{\mathrm{T}}$, $n_{i}$ is the refractive index associated with $\vec{e}_{i}$, and $f(n_{i}) = \mathrm{exp}(\mathrm{i}kn_{i}L)$ is the corresponding complex phasor~\cite{logue2023exploiting}. This field is induced by the input electric field $\vec{\mathcal{E}}_{\mathrm{in}} = \mathcal{E}_{0}\vec{e}_{i}$, whose amplitude is $\mathcal{E}_{0}$ and polarisation is described by the eigenmode $\vec{e}_{i}$. We find an output electric field that depends only on the refractive index associated with the input eigenmode, and therefore the transmission measured after the atomic vapour cell is

\begin{equation}\label{eqn: 2D propagationInputEigenmode2}
\tau_{\vec{e}_{i}} = f^{*}(n_{i})f(n_{i}) = \mathrm{e}^{-2n_{i}^{\mathcal{I}}kL} \,,
\end{equation}

\noindent which again is derived in \ref{App2} (equation~\ref{eqnAppendixB: 2D propagationInputEigenmode3}). Using equations~\ref{eqn: FaradayPropagationEigenmodes} and ~\ref{eqn: VoigtPropagationEigenmodes}, we can write a general expression for the refractive indices of the Faraday and Voigt geometries

\begin{equation}
n_{i, \mathcal{F}/\mathcal{V}} = \sqrt{1 + {\sum}_{q} a_{q}\chi_{q}} \approx 1 + \frac{{\sum}_{q} a_{q}\chi_{q}}{2} \,,
\label{eqn: refractiveIndicesGeneral}
\end{equation}

\noindent where the frequency dependence has been omitted for clarity, and $a_{q}$ are real constants which weight $\chi_{q}$. An approximation of a tenuous atomic medium has also been made ($|\chi_{q}| \ll 1$); this is shown explicitly in both figure~\ref{fig: refractiveIndicesFaraday} and  figure~\ref{fig: refractiveIndicesVoigt}. We can separate equation~\ref{eqn: refractiveIndicesGeneral} into real and imaginary components 

\begin{subequations} \label{eqn: FaradayVoigtGeneralN}
\begin{align}
n_{i, \mathcal{F}/\mathcal{V}}^{\mathcal{R}} - 1 & \approx \frac{{\sum}_{q} a_{q}\chi_{q}^{\mathcal{R}}}{2} \\ n_{i, \mathcal{F}/\mathcal{V}}^{\mathcal{I}} & \approx \frac{{\sum}_{q} a_{q}\chi_{q}^{\mathcal{I}}}{2} = -\frac{\mathrm{ln}(\tau_{\vec{e}_{i}})}{2kL} \,,
\end{align} 
\end{subequations}

\noindent where equation~\ref{eqn: 2D propagationInputEigenmode2} has been used in the final expression. Since the HT is a linear transform

\begin{equation} 
{\sum}_{q}a_{q}\chi_{q}^{\mathcal{R}} = - \mathcal{H}\big[{\sum}_{q}a_{q}\chi_{q}^{\mathcal{I}}\big] \,,
\end{equation}

\noindent therefore the real component of the refractive index is given by

\begin{equation} \label{eqn: FaradayVoigtGeneralRealN}
n_{i, \mathcal{F}/\mathcal{V}}^{\mathcal{R}} \approx 1 -  \mathcal{H}\Big[-\frac{\mathrm{ln}(\tau_{\vec{e}_{i, \mathcal{F}/\mathcal{V}}})}{2kL}\Big] \,.
\end{equation}

In summary, the main difference when applying an external magnetic field is that a specific electric field must be input into an atomic medium in order to
isolate one of its refractive indices. The real component of each index can then be inferred via equation~\ref{eqn: FaradayVoigtGeneralRealN}, which we emphasize is only valid when the eigenmodes of the medium are orthogonal.

\section{Experiment: Measuring the refractive indices of a birefringent atomic medium} \label{sec: experiment}

\begin{figure}[t]
\centering
{\includegraphics[width=1\linewidth]{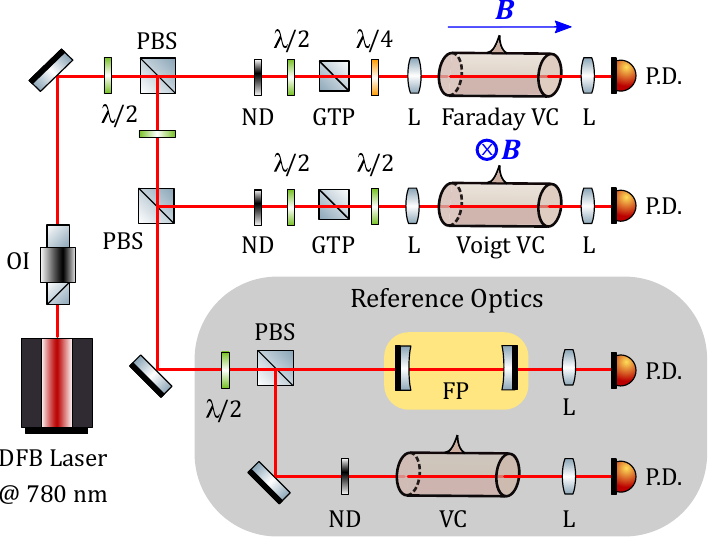}}
\caption{A schematic of the Hilbert transform experiment. Light resonant with the Rb D$_{2}$ line is sourced from a distributed feedback (DFB) diode laser, where its power is split between an experimental and reference arm. Along the experimental arm, a Glan–Taylor polariser (GTP) produces horizontally polarised light, which is then transformed to circular/rotated linear light by a quarter/half-waveplate ($\lambda /4$ \& $\lambda /2$). This light traverses a $2~\mathrm{mm}$ vapour cell (VC) situated in a resistive copper heater, which in the Faraday/Voigt experiment is subject to a longitudinal/transverse magnetic field. The signal is detected using photodetectors (P.D.). Reference optics are used for calibration: a Fabry-P\'{e}rot etalon (FP) removes non-linearities in the laser scan, while a room temperature Rb VC acts as an atomic reference. Power is controlled using pairs of half-waveplates and polarising beam splitter cubes (PBS), in tandem with neutral density filters (ND). Note: for the Faraday experiment only, a second GTP is placed after the VC for one of the measured spectra. OI = optical isolator; M = mirror; L = lens.}

\label{fig: experimentalSetup}
\end{figure}

\begin{figure}[t]
\centering
{\includegraphics[width=0.975\linewidth]{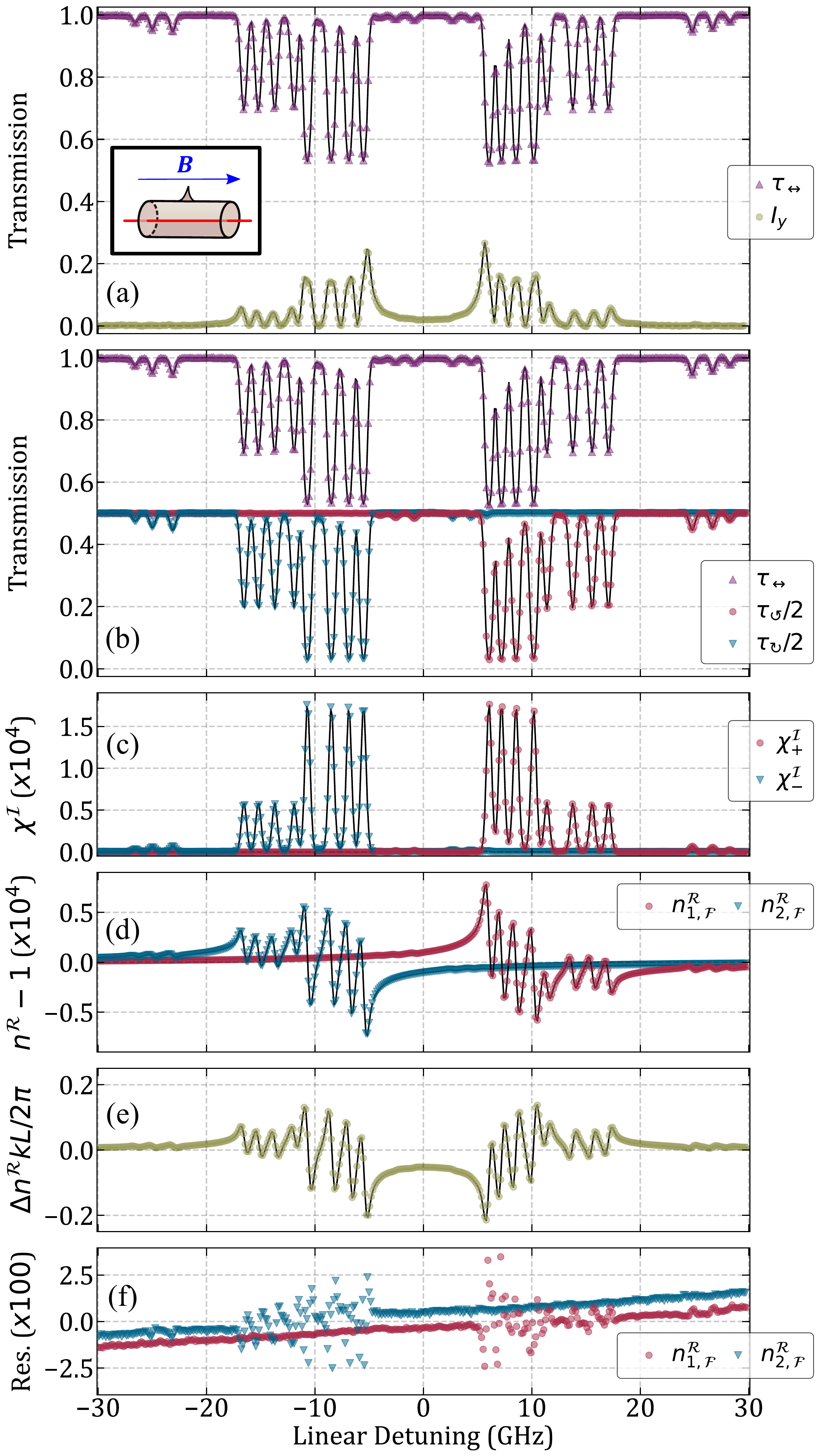}}
\caption{Plot showing application of the Hilbert transform (HT) to indirectly measure the refractive indices of a birefringent atomic medium situated in the Faraday ($\mathcal{F}$) geometry ($\vec{k} \parallel \vec{B}$). Light probes a $^{87}$Rb vapour cell of $\approx 98.2\%$ isotopic abundance in the vicinity of the Rb D$_{2}$ line. The vapour has the following parameters, found by simultaneously fitting the four datasets in panels (a) and (b): $T = (87.02 \pm 0.03)$~$^{\circ}\mathrm{C}$, $B = (5984.0 \pm 0.7)$~$\mathrm{G}$ and $L = 2$~$\mathrm{mm}$. (a) Transmission spectrum $\tau$ for incident horizontal linear light ($\leftrightarrow$, purple) and intensity spectrum $I_{y}$ (olive), found by placing a crossed polariser after the medium aligned with the $y$-axis. (b) Transmission spectra for incident linear horizontal, left-hand circular ($\circlearrowleft$, red), and right-hand circular ($\circlearrowright$, blue) light. (c) Imaginary susceptibilities $\chi^{\mathcal{I}}$ calculated using $\tau_{\circlearrowleft}$ and $\tau_{\circlearrowright}$. (d) Real refractive indices $n^{\mathcal{R}}$ inferred from the imaginary susceptibilities via the HT. The colours used in (c) and (d) are chosen to match (b). (e) Faraday rotation angle, a function of the difference between the inferred refractive indices $\Delta n^{\mathcal{R}}$. This has a one-to-one correspondence with $I_{y}$. For plots (a)-(e), we plot theory curves using $ElecSus$~\cite{zentile2015elecsus, keaveney2018elecsus} and the simultaneous fit parameters.  Since $n^{\mathcal{R}}$ are the crux of the plot, we show their residuals in (f). The residuals are normalised by the maximum value of the corresponding $n^{\mathcal{R}}_{i, \mathcal{F}}$, and multiplied by 100.}
\label{fig: refractiveIndicesFaraday}
\end{figure}

To demonstrate that the HT can be used to indirectly measure the refractive indices of a birefringent atomic medium, we perform an experiment which follows the theory of the previous section. A schematic is shown in figure~\ref{fig: experimentalSetup}. Our setup utilises a distributed feedback laser (DFB) resonant with the Rb D$_{2}$ line, shown to produce a mode-hope free scanning range $>$100$~\mathrm{GHz}$~\cite{ponciano2020absorption}. This light is sent through reference optics: a Rb vapour cell which acts as an atomic reference\footnote{For the Faraday (Voigt) experiment, the atomic reference is $75~\mathrm{mm}$ ($80~\mathrm{mm}$) long and contains $^{85}$Rb with $72.17\%$ ($\approx 100\%$) isotopic purity.}, and a Fabry-P\'{e}rot etalon used to calibrate the frequency axis~\cite{pizzey2022laser}. The light is sent through an experimental arm containing a 2$~\mathrm{mm}$ vapour cell; for the Faraday setup, the cell contains $^{87}$Rb with approximately $98.2\%$ isotopic purity~\cite{higgins2021electromagnetically}, whereas for the Voigt setup, it contains natural abundance Rb. Both cells are heated by a resistive heater and situated in a large magnetic field, orientated in the appropriate geometry. A field strength of $\approx 0.6~\rm{T}$ is achieved using two NdFeB top hat magnets~\cite{briscoe2023voigt}, and therefore the atoms reside in the hyperfine Paschen-Bach regime~\cite{sargsyan2012hyperfine, zentile2014hyperfine, higgins2021electromagnetically, mottola2023electromagnetically}. Before entering the cell, light is linearly polarised in the horizontal direction via the combination of a half-waveplate and Glan-Taylor polariser (GTP). These also control optical power through the cell in tandem with neutral density filters, to order $100~\mathrm{nW}$. We use a beam  waist of size $(373 \pm 2)~\mathrm{\upmu m} \times (559 \pm 2)~\mathrm{\upmu m}$; this ensures the experiment is in the weak probe regime~\cite{sherlock2009weak, siddons2008absolute}, as assumed by our theoretical model. The waist is measured using a Thorlabs CMOS camera and a custom image analysis code, analogous to~\cite{keaveney2018automated}. Light is detected by photodetectors, where it is converted to a voltage and recorded on an oscilloscope. 

For both experiments, we record transmission $\tau$ with a subscript that denotes the input light polarisation. The exception is $I_{y}$, which is the intensity of the output electric field projected onto the $y$-axis of a second GTP, where the initial polarisation is horizontally polarised ($\leftrightarrow$). We would typically refer to this measurement as the atomic filter, since the output polariser is placed after the cell and crossed with respect to the input light polarisation. For the Faraday experiment we record four different outputs, shown in panels (a) and (b) of figure~\ref{fig: refractiveIndicesFaraday}. These are $I_{y}$ (olive) and the transmission spectra of three different input polarisations: linear horizontal ($\tau_{\leftrightarrow}$, purple), left-hand circularly polarised light ($\tau_{\circlearrowleft}$, red) and right-hand circularly polarised light ($\tau_{\circlearrowright}$, blue). Since any linear polarisation can decomposed into equal amounts of left and right handed light~\cite{adams2018optics}, we see that $\tau_{\leftrightarrow} = \tau_{\circlearrowleft}/2 + \tau_{\circlearrowright}/2$. This expression is only true for the Faraday geometry. We record $\tau_{\circlearrowleft, \circlearrowright}$ by placing a quarter waveplate in front of the cell. This produces circularly polarised light, which by equation~\ref{eqn: FaradayPropagationEigenmodes} are eigenmodes of the medium. These two spectra are plotted as  $\tau_{\circlearrowleft, \circlearrowright}/2$ for two reasons: direct comparison with $\tau_{\leftrightarrow}$, and also clarity. A simultaneous fit is used which implements a chi-squared minimisation on the combined residuals of all four outputs. These four outputs form one dataset. In theory, the only variable which is different within each dataset is the measured output transmission. We therefore constrain the fit parameters of each dataset to be the same, with the exception of a small frequency shift ($\approx 10~\mathrm{MHz}$) and a small angle offset ($\approx 1~^{\circ}$) of the second polariser for $I_{y}$. This allows for experimental imperfections in both $x$-axis linearisation and polariser alignment. We use a differential evolution fitting algorithm in tandem with the open-source program $ElecSus$~\cite{zentile2015elecsus, keaveney2018elecsus}, which returns the best fit $T$ and $B$ for that dataset. The variable $T$ denotes the atom temperature, which exponentially scales the number density of an atomic vapour~\cite{pizzey2022laser, alcock1984vapour}. For $I_{y}$ only, the simultaneous fit implies an offset in the second polariser, which we label $\theta_{\mathrm{P}}$. Each dataset is taken four times to account for random errors~\cite{hughes2010measurements}; the global best fit parameters are found by taking the mean $T$, $B$ and $\theta_{\mathrm{P}}$ of all simultaneous fits, while the errors are found by taking the standard error. For the Faraday experiment, we find $T = (87.02 \pm 0.03)$~$^{\circ}\mathrm{C}$, $B = (5984.0 \pm 0.7)$~$\mathrm{G}$ and $\theta_{\mathrm{P}} = (1.12 \pm 0.03)$~$^{\circ}$. As shown in figure~\ref{fig: refractiveIndicesFaraday}, excellent agreement between experiment and theory can be seen\footnote{We note that the errors obtained from fitting the datasets are unreasonably small, and a more realistic measure is $T = (87.0 \pm 0.7)$~$^{\circ}\mathrm{C}$ and $B = (5984 \pm 9)$~$\mathrm{G}$. For $T$, we estimate the temperature offset assuming an approximate $5~\%$ error in the atomic number density equation used by $ElecSus$~\cite{zentile2015elecsus, alcock1984vapour}, and for $B$, we convert the average error on the atomic reference frequency shift/calibration to a corresponding magnetic field (see~\cite{zentile2014hyperfine} for a similar discussion). Very similar errors are found for the Voigt experiment, since it utilises the same reference optics.}.

The rest of figure~\ref{fig: refractiveIndicesFaraday} follows the methodology shown in figure~\ref{fig: HilbertMethod} to determine the real refractive indices: panel (c) shows the $\pm 1$ components of $\chi^{\mathcal{I}}$, found using equation~\ref{eqn: 2D propagationInputEigenmode2} after substitution of the Faraday refractive indices in equation~\ref{eqn: FaradayPropagationEigenmodes}. These signals are then Hilbert transformed, such that the real components of each refractive index can be inferred in panel (d). We plot the real component of the refractive indices returned by $ElecSus$ using the simultaneous fit parameters; from the residuals in (f), we see excellent agreement between theory and the Hilbert transformed data. Only at the outer edges are systematic deviations seen, caused by using a truncated dataset which does not span $\pm\infty$, like the HT and KK relations do. We note that the total size of laser scan used in the Faraday experiment is $\approx 110~\mathrm{GHz}$, and that the HT is taken over the whole scan range; figure~\ref{fig: refractiveIndicesFaraday} has been cropped to only show the main region of atom-light interaction. If we instead take the HT on the $\pm 30~\mathrm{GHz}$ range shown, the root-mean-square (RMS) of the residuals in (f) increases from $0.8\%$ ($0.9\%$) to $2.5\%$ ($2.7\%$) for $n_{1, \mathcal{F}}$ ($n_{2, \mathcal{F}}$). Even at a lower frequency range, the transformed data still shows very good agreement with theory. Panel (e) shows the Faraday rotation $\theta_{\mathcal{F}} = (n^{\mathcal{R}}_{2, \mathcal{F}} - n^{\mathcal{R}}_{1, \mathcal{F}})kL/2$ in units of $\pi$~\cite{kemp2011analytical, weller2012measuring, zentile2014hyperfine}, which is a function of the difference between the two refractive indices in panel (d). The HT technique therefore provides an indirect measurement of atomic magneto-optical rotation. By comparison of panel (e) with $I_{y}$ in panel (a), we see that atomic birefringence, and therefore magneto-optical rotation, is the foundation of the atomic filter. 

\begin{figure}[t]
\centering
{\includegraphics[width=1.0\linewidth]{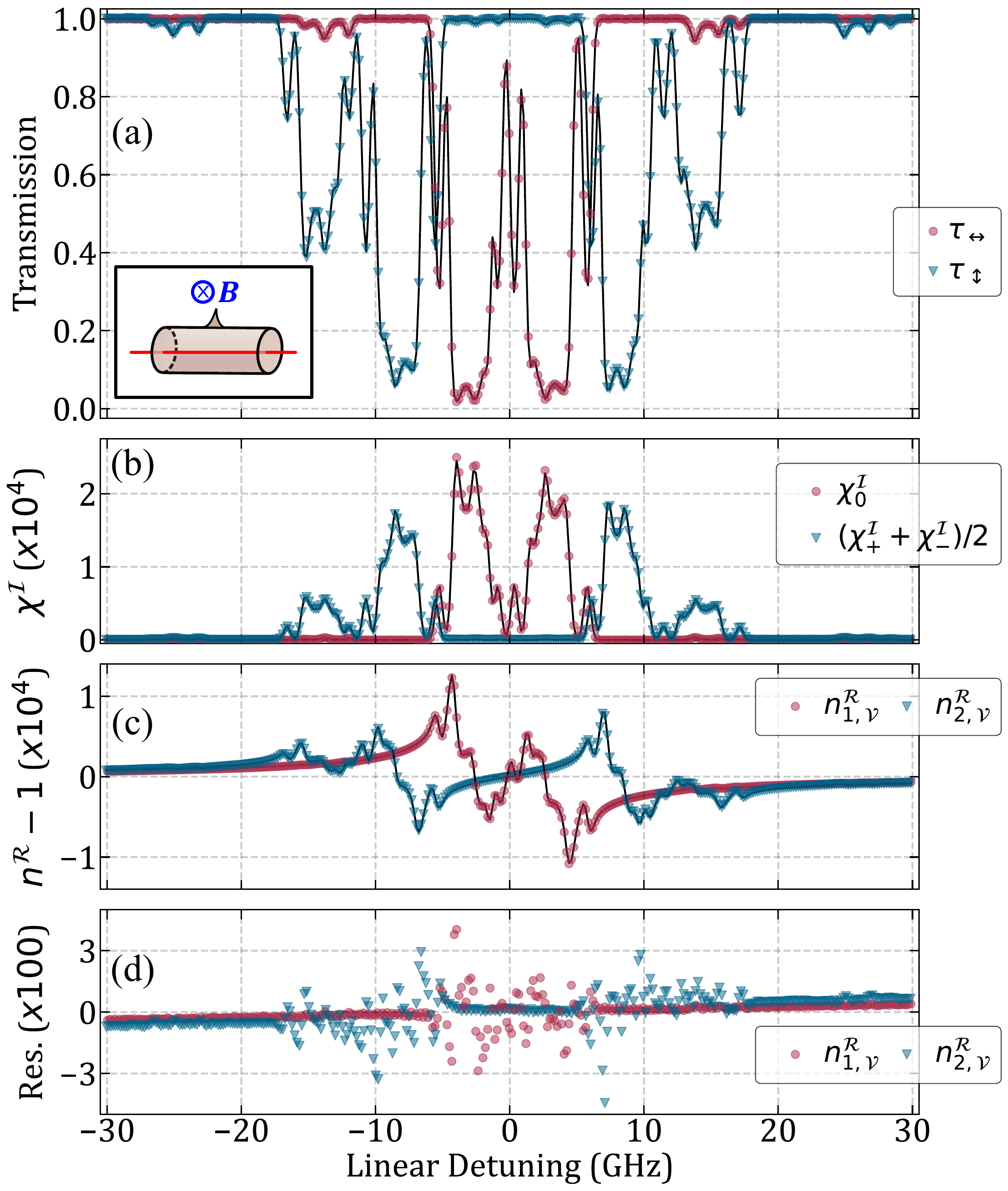}}
\caption{Plot showing application of the Hilbert transform (HT) to indirectly measure the refractive indices of a birefringent atomic medium situated in the Voigt ($\mathcal{V}$) geometry ($\vec{k} \perp \vec{B}$). Light resonant with the Rb D$_{2}$ line probes a Rb vapour cell of natural abundance. The vapour has the following parameters, found by simultaneously fitting the two datasets in panel (a): $T = (99.31 \pm 0.01)~^{\circ}\rm{C}$, $B = (6006.0 \pm 0.4)~\mathrm{G}$ and $L = 2$~$\mathrm{mm}$. (a) Transmission spectrum $\tau$ for incident horizontal linear ($\leftrightarrow$, red) and vertical linear ($\updownarrow$, blue) light. (b) Imaginary susceptibilities $\chi^{\mathcal{I}}$ calculated using $\tau_{\leftrightarrow}$ and $\tau_{\updownarrow}$. (c) Real refractive indices $n^{\mathcal{R}}$ inferred from the imaginary susceptibilities via the HT, and (d), residuals for $n^{\mathcal{R}}$. The colours used in (b)-(d) are chosen to match (a). Theory curves are found using $ElecSus$~\cite{zentile2015elecsus, keaveney2018elecsus} and the simultaneous fit parameters. The residuals are normalised by the maximum value of the corresponding $n^{\mathcal{R}}_{i, \mathcal{V}}$, and multiplied by 100.}
\label{fig: refractiveIndicesVoigt}
\end{figure}

For the Voigt experiment, we record the two outputs needed to measure $n^{\mathcal{R}}_{i, \mathcal{V}}$. These are shown in panel (a) of figure~\ref{fig: refractiveIndicesVoigt}, and are transmission spectra for the input polarisations linear horizontal ($\tau_{\leftrightarrow}$, red) and linear vertical ($\tau_{\updownarrow}$, blue). The experimental arm uses a half waveplate in front of the cell to rotate the input plane of polarisation, producing the eigenmodes of Voigt geometry (equation~\ref{eqn: VoigtPropagationEigenmodes}). We use the same fitting method, returning $T = (99.31 \pm 0.01)~^{\circ}\rm{C}$ and $B = (6006.0 \pm 0.4)~\mathrm{G}$. Both refractive indices are indirectly measured, following the procedure outlined in figure~\ref{fig: HilbertMethod}. Excellent agreement between theory and experiment can again be seen. For the Voigt experiment, we used a much larger scan range of $\approx 210~\mathrm{GHz}$; taking the HT on the $\pm 30~\mathrm{GHz}$ range shown increases the RMS of the residuals in (d) from $0.6\%$ ($0.9\%$) to $3.2\%$ ($5.3\%$) for $n_{1, \mathcal{V}}$ ($n_{2, \mathcal{V}}$). We find that the expected RMS increase from using natural abundance Rb, which has a significantly more complex atomic structure, is offset by the RMS decrease from using a larger scan range (compared to the Faraday experiment). This emphasises the utility of HT to measuring the refractive indices of any atomic system in both the Faraday and Voigt geometries. We add that, as discussed in~\cite{whittaker2015hilbert}, the method is limited to a regime which is not optically thick. However, we do expect the method to work at all magnetic fields, with $0.6~\mathrm{T}$ chosen in this study to highlight the complex resonance structure of atomic media. From both figures 3 and 4, we clearly see that the HT encapsulates the full dispersive behaviour of both systems.

\section{Conclusion} \label{sec: conclusion}
In this study, we have utilised the HT as a method of  measuring both refractive indices of a birefringent atomic medium. We have shown the method is valid in two cases which differ by the relative direction of the magnetic field and light wavevector; these are commonly referred to as the Faraday and Voigt geometries. In these geometries, the light propagation algebra is analytic, and therefore an atom-light interaction that is associated with only one of the two refractive indices can be induced. We showed that this interaction is caused by specifically inputting a light polarisation into the atomic medium which matches one of its propagation eigenmodes. The output absorption spectrum is therefore only a function of the refractive index which corresponds to the input eigenmode. We can then measure the imaginary component of each refractive index $n^{\mathcal{I}}_{i}$ via absorption spectroscopy, so that each real component $n^{\mathcal{R}}_{i}$ can be inferred via the HT. This is significant, as typically $n^{\mathcal{R}}$ is very difficult to measure, and $n^{\mathcal{I}}$ is not. Since both refractive indices can be measured independently, the method provides an indirect measure of atomic magneto-optical rotation. This work builds upon the work presented in~\cite{whittaker2015hilbert}, and broadens the range of applications of the HT in the field of linear optics and atomic physics when an atomic medium is subject to an external magnetic field. 

\section*{Acknowledgements}
The authors would like to thank Fraser Logue for fruitful discussions, Liam A. P. Gallagher for proof-reading the manuscript, and Alex Matthies for helping with figure visualisation.

\section*{Funding}
The authors thank EPSRC for funding this work (Grant No. EP/T518001/1 and EP/R002061/1). 

\section*{Submission statement}
This is the version of the article before peer review or editing, as submitted by an author to Journal of Physics B. IOP Publishing Ltd is not responsible for any errors or omissions in this version of the manuscript or any version derived from it. If accepted, the arXiv submission will be updated to include the journal reference and related DOI.

\section*{Disclosures}
The authors declare no conflicts of interest.

\section*{Data availability}
Data underlying the results presented in this paper are available from~\cite{data}.

\appendix
\section{Propagating light through an atomic thermal vapour} \label{App1} 
The text below summarises work presented in~\cite{rotondaro2015generalized, keaveney2018elecsus}, and is used in \ref{App2}. The output electric field $\vec{\mathcal{E}}_{\mathrm{out}}$ after light traverses an atomic medium is given by

\begin{equation}
\mathcal{\vec{E}}_{\mathrm{out}} = R^{-1} P R \mathcal{\vec{E}}_{\mathrm{in}} \,,
\label{eqnAppendix: propagationEout}
\end{equation}

\noindent where $\vec{\mathcal{E}}_{\mathrm{in}} = (\mathcal{E}_{x}, \mathcal{E}_{y}, \mathcal{E}_{z})^{\mathrm{T}}$ is the input electric field vector, $R$ is a rotation matrix, and $P$ is the propagation matrix. Both $R$ and $P$ are derived using the dispersion equation~(\ref{eqnAppendix: dispersionRelation}) for a non-magnetic dielectric medium subject to plane monochromatic radiation~\cite{rotondaro2015generalized, keaveney2018elecsus, palik1970infrared}

\begin{equation}
\begin{pmatrix}
(\epsilon_{xx} - n^{2})\,\mathrm{cos}\,\theta_{B} & \epsilon_{xy} & \epsilon_{xx}\,\mathrm{sin}\,\theta_{B}\\
-\epsilon_{xy}\,\mathrm{cos}\,\theta_{B} & \epsilon_{xx} - n^{2} & -\epsilon_{xy}\,\mathrm{sin}\,\theta_{B}\\
-(\epsilon_{zz} - n^{2})\,\mathrm{sin}\,\theta_{B} & 0 & \epsilon_{zz}\,\mathrm{cos}\,\theta_{B}\\
\end{pmatrix}
\begin{pmatrix}
\mathcal{E}_{x}\\
\mathcal{E}_{y}\\
0\\
\end{pmatrix}
= 0 \,,
\label{eqnAppendix: dispersionRelation}
\end{equation}

\noindent where $\theta_{B}$ is the angle of the magnetic field vector $\vec{B}$ relative to the light wavevector $\vec{k}$, and $\epsilon_{ij}$ are components of the dielectric tensor $\vec{\epsilon}$~\cite{rotondaro2015generalized}

\begin{subequations}\label{eqnAppendix: dielectricTensorComponents}
\begin{align}
\epsilon_{xx} & = 1 + (\chi_{\mathrm{+1}} + \chi_{\mathrm{-1}})/2 \\
\epsilon_{xy} & = \mathrm{i}(\chi_{\mathrm{-1}} - \chi_{\mathrm{+1}})/2 \\
\epsilon_{zz} & = 1 + \chi_{\mathrm{0}} \,.
\end{align}
\end{subequations}

\noindent As described below, equations~\ref{eqnAppendix: dielectricTensorComponents} therefore link the macroscopic refractive indices $n$ of an atomic medium to its microscopic electric susceptibility $\chi$~\cite{keaveney2018elecsus, zentile2015elecsus}. We associate components of $\chi$ with electric-dipole transition selection rules: a transition that induces a change in an atomic state's projection quantum number $\Delta m_{l} = \pm 1$, where $l$ is the orbital angular momenta, is known as a $\sigma^{\pm}$ transition respectively, whereas light which induces $\Delta m_{l} = 0$ drives $\pi$ transitions~\cite{foot2004atomic}. The transition profiles for each atomic resonance are summed separately for $\Delta m_{l} \in \{+1, -1, 0\}$, and assigned to the matching component of $\chi \in \{\chi_{+1}, \chi_{-1}, \chi_{0}\}$. These are then incorporated into the model via the dispersion equation. A detailed discussion of how our theoretical model calculates atomic state eigenenergies, and therefore transition energies, can be found in~\cite{wellerThesis, zentile2015elecsus}. The shape of the transition profile for each resonance is discussed in~\cite{siddons2008absolute, pizzey2022laser}. We note that equation~\ref{eqnAppendix: dispersionRelation} assumes an input transverse electric field whose polarisation is constrained to the $x$-$y$ plane. The field therefore propagates along the $z$-axis ($\vec{k} = k\hat{z}$), and consequently, the $z$ component of the electric field polarisation equals zero ($\mathcal{E}_{z} =~0$). By setting the determinant of the $3 \times 3$ matrix in equation~\ref{eqnAppendix: dispersionRelation} to zero, we solve for the two complex refractive indices $n_{1,2}$ of a birefringent and dichroic atomic medium. The refractive indices are functions $\epsilon_{ij}$, and therefore the different components of $\chi$. Each refractive index $n_{i}$ is substituted back into equation~\ref{eqnAppendix: dispersionRelation} to find its corresponding propagation eigenmode $\vec{e}_{i} = (e_{i1}, e_{i2}, e_{i3})^{\mathrm{T}}$

\begin{equation}
\begin{pmatrix}
(\epsilon_{xx} - n^{2}_{i})\,\mathrm{cos}\,\theta_{B} & \epsilon_{xy} & \epsilon_{xx}\,\mathrm{sin}\,\theta_{B}\\
-\epsilon_{xy}\,\mathrm{cos}\,\theta_{B} & \epsilon_{xx} - n^{2}_{i} & -\epsilon_{xy}\,\mathrm{sin}\,\theta_{B}\\
-(\epsilon_{zz} - n^{2}_{i})\,\mathrm{sin}\,\theta_{B} & 0 & \epsilon_{zz}\,\mathrm{cos}\,\theta_{B}\\
\end{pmatrix}
\begin{pmatrix}
e_{i1} \\
e_{i2} \\
e_{i3} \\
\end{pmatrix}
= 0 \,.
\label{eqnAppendix: dispersionRelationSolve}
\end{equation}

\noindent We use the components of $\vec{e}_{i}$ to populate the rotation matrix 

\begin{equation}
R = 
\begin{pmatrix}
e_{11} & e_{12} & e_{13} \\
e_{21} & e_{22} & e_{23} \\
0 & 0 & 1 \\
\end{pmatrix}^{*} \,,
\label{eqnAppendix: rotationMatrix}
\end{equation}

\noindent where the first and second rows are the eigenmodes of $n_{1,2}$ respectively, and * represents the complex conjugate. Using equation~\ref{eqnAppendix: propagationEout}, we see that applying $R$ to $\mathcal{\vec{E}}_{\mathrm{in}}$ from the left projects the polarisation of input light onto the eigenbasis of the atom-light system~\cite{rotondaro2015generalized}. The light is then propagated a distance $z = L$ along the $z$-axis via application of the translation matrix $P$

\begin{equation}
P = 
\begin{pmatrix}
e^{\mathrm{i}kn_{1}L} & 0 & 0 \\
0 & e^{\mathrm{i}kn_{2}L} & 0 \\
0 & 0 & 1 \\
\end{pmatrix} 
=
\begin{pmatrix}
f(n_{1}) & 0 & 0 \\
0 & f(n_{2}) & 0 \\
0 & 0 & 1 \\
\end{pmatrix} \,,
\label{eqnAppendix: translationMatrix}
\end{equation}

\noindent where each element $f(n_{i}) = \mathrm{exp}(\mathrm{i}kn_{i}L)$ is a complex phasor associated with the eigenmode $\vec{e}_{i}$~\cite{logue2023exploiting}, and $L$ is the length of the atomic medium. The propagation matrix is diagonal since each row of $R\mathcal{\vec{E}}_{\mathrm{in}}$ is the input Cartesian polarisation projected onto an eigenmode of the atom-light coordinate system, with each row corresponding to a separate eigenmode. Since light is transverse, only the $x$-$y$ components of $\mathcal{\vec{E}}_{\mathrm{in}}$ evolve, and therefore the third diagonal component of $P$ is $1$. The electric field is mapped back to Cartesian coordinates via the inverse rotation matrix $R^{-1}$.

The dispersion equation can be solved analytically for the Faraday ($\theta_{B} = 0$) and Voigt ($\theta_{B} = \pi/2$) cases using the method described above. Both their refractive indices and corresponding propagation eigenmodes can also be found in literature~\cite{rotondaro2015generalized, logue2023exploiting, keaveney2018elecsus, palik1970infrared}. The Faraday ($\mathcal{F}$) solutions are 

\begin{subequations} \label{eqnAppendix: FaradayPropagationEigenmodes}
\begin{align}
& \left.\begin{array}{ll}
    n_{\mathrm{1, \mathcal{F}}} & = \sqrt{\epsilon_{xx} + \mathrm{i}\epsilon_{xy}}  \\
    & \approx 1 + \chi_{\mathrm{+}}/2 \\
    \end{array}\right\} \quad \vec{e}_{1, \mathcal{F}} = \frac{1}{\sqrt{2}}\begin{pmatrix} 1 & \mathrm{i} 
    \end{pmatrix}^{\mathrm{T}} \\
& \left.\begin{array}{ll}
    n_{\mathrm{2}, \mathcal{F}} & = \sqrt{\epsilon_{xx} - \mathrm{i}\epsilon_{xy}}  \\
    & \approx 1 + \chi_{\mathrm{-}}/2 \\
    \end{array}\right\} \quad \vec{e}_{2, \mathcal{F}} = \frac{1}{\sqrt{2}}\begin{pmatrix} 1 & -\mathrm{i} 
    \end{pmatrix}^{\mathrm{T}} \,,
\end{align}
\end{subequations}

\noindent and the Voigt ($\mathcal{V}$) solutions are

\begin{subequations} \label{eqnAppendix: VoigtPropagationEigenmodes}
\begin{align}
& \left.\begin{array}{ll}
    n_{\mathrm{1, \mathcal{V}}} & = \sqrt{\epsilon_{zz}} \\
    & \approx 1 + \chi_{\mathrm{0}}/2 \\
    \end{array} \quad\quad\quad  \right\} \quad \vec{e}_{1, \mathcal{V}} = \begin{pmatrix} 1 & 0 
    \end{pmatrix}^{\mathrm{T}} \\
& \left.\begin{array}{ll}
    n_{\mathrm{2}, \mathcal{V}} & = \sqrt{\epsilon_{xx} + \epsilon_{xy}^{2}/\epsilon_{xx}}  \\
    & \approx 1 + (\chi_{\mathrm{+}} + \chi_{\mathrm{-}})/4 \\
    \end{array}\right\} \quad \vec{e}_{2, \mathcal{V}} = \begin{pmatrix} 0 &  1 
    \end{pmatrix}^{\mathrm{T}} \,.
\end{align}
\end{subequations}

\noindent In all cases, we have used a tenuous atomic medium approximation i.e. $|\chi| \ll 1$, and each eigenmode has been normalised. Following~\cite{logue2023exploiting}, we can use the transverse nature of the electric field to reduce the system to two dimensions. This means that only the upper left $2 \times 2$ minor of the matrices in equations~\ref{eqnAppendix: rotationMatrix} and~\ref{eqnAppendix: translationMatrix} are relevant in propagation calculations.

\section{Isolating the refractive indices of a birefringent atomic thermal vapour} \label{App2} 

This appendix details a method to isolate the refractive indices of a birefringent atomic thermal vapour. For any atom-light system described by equation~\ref{eqnAppendix: dispersionRelation}, we prove that this is possible by inputting an electric field into an atomic medium which is one of its propagation eigenmodes. The proof requires one condition: the system must exhibit orthogonal eigenmodes. This is true for Faraday and Voigt (equations~\ref{eqnAppendix: FaradayPropagationEigenmodes} and ~\ref{eqnAppendix: VoigtPropagationEigenmodes}), but not in general~\cite{logue2023exploiting}.

To start, we follow \ref{App1} in order to derive a general equation for the output electric field vector after light traverses an atomic medium. Using equations \ref{eqnAppendix: propagationEout}, \ref{eqnAppendix: rotationMatrix} and \ref{eqnAppendix: translationMatrix}

\begin{align}\label{eqnAppendixB: 2D propagation}
& \vec{\mathcal{E}}_{\mathrm{out}} = R^{-1}TR\vec{\mathcal{E}}_{\mathrm{in}} 
 = \frac{1}{e_{11}^{*}e_{22}^{*} - e_{12}^{*}e_{21}^{*}} \times \\ & 
\begin{pmatrix}
 e_{22}^{*}f(n_{1})[\mathcal{E}_{1}e_{11}^{*} + \mathcal{E}_{2}e_{12}^{*}]  - e_{12}^{*}f(n_{2})[\mathcal{E}_{1}e_{21}^{*} + \mathcal{E}_{2}e_{22}^{*}] \\
e_{11}^{*}f(n_{2})[\mathcal{E}_{1}e_{21}^{*} + \mathcal{E}_{2}e_{22}^{*}] -e_{21}^{*}f(n_{1})[\mathcal{E}_{1}e_{11}^{*} + \mathcal{E}_{2}e_{12}^{*}] \nonumber
\end{pmatrix}  \,,
\end{align}

\noindent where $n_{i}$ is the refractive index associated with the eigenmode $\vec{e}_{i} = (e_{i1}, e_{i2})^{\mathrm{T}}$, and the input transverse light is described by the polarisation vector $\vec{\mathcal{E}}_{\mathrm{in}} = (\mathcal{E}_{1}, \mathcal{E}_{2})^{\mathrm{T}}$. Note there is no $z$-component, since the system has been reduced to two-dimensions\footnote{Using a full three dimensional expansion of equation~\ref{eqnAppendix: propagationEout}, we find the $z$ component of the output electric field vector $\mathcal{E}_{\mathrm{out, z}} = 0$, since the $z$ component of the input electric field vector $\mathcal{E}_{\mathrm{in, z}} = 0$.}. There are two cases to consider, each corresponding to one of the eigenmodes of the system. Explicitly, these are $\vec{\mathcal{E}}_{\mathrm{in}} = \mathcal{E}_{0}\vec{e}_{1/2}$, where $\mathcal{E}_{0}$ is the amplitude of the electric field. In this appendix, the proof is only shown for $\vec{e}_{1}$, as it is trivial to extend the proof to $\vec{e}_{2}$. We input into the medium the vector $\vec{\mathcal{E}}_{\mathrm{in}} = \mathcal{E}_{0}\vec{e}_{1} = (\mathcal{E}_{0}e_{11}, \mathcal{E}_{0}e_{12})^{\mathrm{T}}$ i.e. $\mathcal{E}_{1}/\mathcal{E}_{0} = e_{11}$ and $\mathcal{E}_{2}/\mathcal{E}_{0} = e_{12}$, and find

\begin{align}\label{eqnAppendixB: 2D propagationInputEigenmode1}
&  R^{-1}TR\mathcal{E}_{0}\vec{e}_{1} = \frac{\mathcal{E}_{0}}{e_{11}^{*}e_{22}^{*} - e_{12}^{*}e_{21}^{*}} \times \\ &
\begin{pmatrix}
e_{22}^{*}f(n_{1})[\lvert e_{11} \rvert^{2} + \lvert e_{12} \rvert^{2}] - e_{12}^{*}f(n_{2})[e_{11}e_{21}^{*} + e_{12}e_{22}^{*}] \\
e_{11}^{*}f(n_{2})[e_{11}e_{21}^{*} + e_{12}e_{22}^{*}] -  e_{21}^{*}f(n_{1})[\lvert e_{11} \rvert^{2} + \lvert e_{12} \rvert^{2}]  \nonumber
\end{pmatrix}  \,,
\end{align}


\noindent where $\lvert e_{ij} \rvert^{2} = e_{ij}e_{ij}^{*}$. Next, we make the assumption that the eigenmodes of an atomic medium are orthogonal. This is only true for the Faraday and Voigt geometries~\cite{logue2023exploiting}. In this regime, $\vec{e}_{1}\cdot\vec{e}_{2} = 0$, and therefore

\begin{equation}\label{eqnAppendixB: orthogonalEigenmodeSimplification}
e_{11}^{*}e_{21} = -e_{12}^{*}e_{22} \quad \equiv \quad e_{11}e_{21}^{*} = -e_{12}e_{22}^{*} \,.
\end{equation}

\noindent Substituting equation~\ref{eqnAppendixB: orthogonalEigenmodeSimplification} into equation~\ref{eqnAppendixB: 2D propagationInputEigenmode1}

\begin{equation}\label{eqnAppendixB: 2D propagationInputEigenmode2}
\begin{split}
\vec{\mathcal{E}}_{\mathrm{out}} & = R^{-1}TR\mathcal{E}_{0}\vec{e}_{1} = f(n_{1})\mathcal{E}_{0}  \begin{pmatrix}
e_{11} \\
e_{12} \\
\end{pmatrix} \\
& = f(n_{1})\mathcal{E}_{0}\vec{e}_{1} \,,
\end{split}
\end{equation}

\noindent  which depends only on the refractive index associated with the input propagation eigenmode. The transmission measured by a photodetector after an atomic vapour cell is therefore 

\begin{equation}\label{eqnAppendixB: 2D propagationInputEigenmode3}
\begin{split}
\tau_{\vec{e}_{1}} & = \frac{\vec{\mathcal{E}}_{\mathrm{out}}^{\dag}\vec{\mathcal{E}}_{\mathrm{out}}}{\vec{\mathcal{E}}_{\mathrm{in}}^{\dag}\vec{\mathcal{E}}_{\mathrm{in}}} \\
& = f^{*}(n_{1})f(n_{1}) = \mathrm{e}^{-2n_{1}^{\mathcal{I}}kL} \,,
\end{split}
\end{equation}

\noindent where $n_{1} = n_{1}^{\mathcal{R}} + \mathrm{i}n_{1}^{\mathcal{I}}$ is used, and the subscript $\vec{e}_{1}$ denotes using a propagation eigenmode as the input light vector. By rearranging equation~\ref{eqnAppendixB: 2D propagationInputEigenmode3}, the imaginary component of the complex refractive index $n$ can be found and used to determine the real component via the HT. We therefore show how one can isolate the two refractive indices of a birefringent atomic thermal vapour under the assumption of orthogonal propagation eigenmodes.

\section*{References}
\bibliographystyle{iopart-num}
\bibliography{bib.bib}

\providecommand{\newblock}{}
\begin{thebibliography}{100}
\expandafter\ifx\csname url\endcsname\relax
  \def\url#1{{\tt #1}}\fi
\expandafter\ifx\csname urlprefix\endcsname\relax\def\urlprefix{URL }\fi
\providecommand{\eprint}[2][]{\url{#2}}

\bibitem{lovell1974application}
Lovell R 1974 {\em J. Phys. C: Solid State Phys.\/} {\bf 7} 4378

\bibitem{roessler1965kramers}
Roessler D~M 1965 {\em Br. J. Appl. Phys.\/} {\bf 16} 1119

\bibitem{bakry2018using}
Bakry M and Klinkenbusch L 2018 {\em ARS\/} {\bf 16} 23--28

\bibitem{horsley2015spatial}
Horsley S~A~R, Artoni M and La~Rocca G~C 2015 {\em Nat. Photonics\/} {\bf 9} 436--439

\bibitem{sai2020designing}
Sai T, Saba M, Dufresne E~R, Steiner U and Wilts B~D 2020 {\em Faraday Discuss.\/} {\bf 223} 136--144

\bibitem{darwish2019deposition}
Darwish A~A~A, Aboraia A~M, Soldatov A~V and Yahia I~S 2019 {\em Opt. Mater.\/} {\bf 95} 109219

\bibitem{o1981kramers}
O’Donnell M, Jaynes E~T and Miller J~G 1981 {\em J. Acoust. Soc. Am.\/} {\bf 69} 696--701

\bibitem{szabo2010unique}
Szab{\'o} Z, Park G~H, Hedge R and Li E~P 2010 {\em IEEE Trans. Microw. Theory Tech.\/} {\bf 58} 2646--2653

\bibitem{macdonald1985application}
Macdonald D~D and Urquidi-Macdonald M 1985 {\em J. Electrochem. Soc.\/} {\bf 132} 2316

\bibitem{tanner2015use}
Tanner D~B 2015 {\em Phys. Rev. B\/} {\bf 91} 035123

\bibitem{mecozzi2016kramers}
Mecozzi A, Antonelli C and Shtaif M 2016 {\em Optica\/} {\bf 3} 1220--1227

\bibitem{harter2020generalized}
Harter T, F{\"u}llner C, Kemal J~N, Ummethala S, Steinmann J~L, Brosi M, Hesler J~L, Br{\"u}ndermann E, M{\"u}ller A~S, Freude W, Randel S and Koos C 2020 {\em Nat. Photonics\/} {\bf 14} 601--606

\bibitem{faber2004oxygen}
Faber D~J, Aalders M~C~G, Mik E~G, Hooper B~A, van Gemert M~J~C and van Leeuwen T~G 2004 {\em Phys. Rev. Lett.\/} {\bf 93} 028102

\bibitem{baek2021intensity}
Baek Y and Park Y 2021 {\em Nat. Photonics\/} {\bf 15} 354--360

\bibitem{huang2022high}
Huang Z and Cao L 2022 {\em Adv. Photonics Res.\/} {\bf 3} 2100273

\bibitem{toll1956causality}
Toll J~S 1956 {\em Phys. Rev.\/} {\bf 104} 1760

\bibitem{kronig1926theory}
Kronig R~d~L 1926 {\em J. Opt. Soc. Am.\/} {\bf 12} 547--557

\bibitem{kramers1927diffusion}
Kramers H~A 1927 {\em Atti Congr. Internaz. Fisici (Transactions of Volta Centenary Congress), Como\/} {\bf 2} 545--557

\bibitem{demtroder1982laser}
Demtr{\"o}der W 1982 {\em {Laser spectroscopy}\/} vol~1 (Springer)

\bibitem{adams2018optics}
Adams C~S and Hughes I~G 2018 {\em {Optics f2f: from Fourier to Fresnel}\/} (Oxford University Press)

\bibitem{camacho2007wide}
Camacho R~M, Pack M~V, Howell J~C, Schweinsberg A and Boyd R~W 2007 {\em Phys. Rev. Lett.\/} {\bf 98} 153601

\bibitem{siddons2008absolute}
Siddons P, Adams C~S, Ge C and Hughes I~G 2008 {\em J. Phys. B: At. Mol. Opt. Phys.\/} {\bf 41} 155004

\bibitem{pizzey2022laser}
Pizzey D, Briscoe J~D, Logue F~D, Ponciano-Ojeda F~S, Wrathmall S~A and Hughes I~G 2022 {\em New J. Phys.\/} {\bf 24} 125001

\bibitem{purves2004refractive}
Purves G~T, Jundt G, Adams C~S and Hughes I~G 2004 {\em Eur. Phys. J. D\/} {\bf 29} 433--436

\bibitem{xiao1995measurement}
Xiao M, Li Y~Q, Jin S~Z and Gea-Banacloche J 1995 {\em Phys. Rev. Lett.\/} {\bf 74} 666

\bibitem{keaveney2012maximal}
Keaveney J, Hughes I~G, Sargsyan A, Sarkisyan D and Adams C~S 2012 {\em Phys. Rev. Lett.\/} {\bf 109} 233001

\bibitem{stavenga2013quantifying}
Stavenga D~G, Leertouwer H~L and Wilts B~D 2013 {\em Light Sci. Appl.\/} {\bf 2} Article E100

\bibitem{whittaker2015hilbert}
Whittaker K~A, Keaveney J, Hughes I~G and Adams C~S 2015 {\em Phys. Rev. A\/} {\bf 91} 032513

\bibitem{king_2009}
King F~W 2009 {\em {Hilbert Transforms}\/} ({\em Encyclopedia of Mathematics and its Applications\/} vol~1) (Cambridge University Press)

\bibitem{boyd2020}
Boyd R~W 2020 {\em {Nonlinear Optics}\/} 4th ed (Elsevier, Academic Press)

\bibitem{benitez2001use}
Benitez D, Gaydecki P~A, Zaidi A and Fitzpatrick A~P 2001 {\em Comput. Biol. Med.\/} {\bf 31} 399--406

\bibitem{volkov2020situ}
Volkov O, Pavlovskiy V, Gundareva I, Khabibullin R and Divin Y 2020 {\em IEEE Trans. Terahertz Sci.\/} {\bf 11} 330--338

\bibitem{li2016phase}
Li W~S, Chen C~W, Lin K~F, Chen H~R, Tsai C~Y, Chen C~H and Hsieh W~F 2016 {\em Opt. Lett,\/} {\bf 41} 1616--1619

\bibitem{zhu2012electron}
Zhu X, Santos L, Howard C, Sankar R, Chou F~C, Chamon C and El-Batanouny M 2012 {\em Phys. Rev. Lett.\/} {\bf 108} 185501

\bibitem{davis2000image}
Davis J~A, McNamara D~E, Cottrell D~M and Campos J 2000 {\em Opt. Lett.\/} {\bf 25} 99--101

\bibitem{mishnev1993discrete}
Mishnev A~F 1993 {\em Acta Cryst. A\/} {\bf 49} 159--161

\bibitem{feldman2011hilbert}
Feldman M 2011 {\em MSSP\/} {\bf 25} 735--802

\bibitem{virtanen2020scipy}
Virtanen P {\em et~al.\/} 2020 {\em Nat. Methods\/} {\bf 17} 261--272

\bibitem{lucarini2005kramers}
Lucarini V, Saarinen J~J, Peiponen K~E and Vartiainen E~M 2005 {\em {Kramers-Kronig relations in optical materials research}\/} (Springer Science \& Business Media)

\bibitem{breit1931measurement}
Breit G and Rabi I~I 1931 {\em Phys. Rev.\/} {\bf 38} 2082

\bibitem{tremblay1990absorption}
Tremblay P, Michaud A, Levesque M, Th{\'e}riault S, Breton M, Beaubien J and Cyr N 1990 {\em Phys. Rev. A\/} {\bf 42} 2766

\bibitem{umfer1992investigations}
Umfer C, Windholz L and Musso M 1992 {\em Z. Phys. D\/} {\bf 25} 23--29

\bibitem{windholz1985zeeman}
Windholz L 1985 {\em Z. Phys. A\/} {\bf 322} 203--206

\bibitem{windholz1988zeeman}
Windholz L and Musso M 1988 {\em Z. Phys. D\/} {\bf 8} 239--249

\bibitem{ponciano2020absorption}
Ponciano-Ojeda F~S, Logue F~D and Hughes I~G 2020 {\em J. Phys. B: At. Mol. Opt. Phys.\/} {\bf 54} 015401

\bibitem{pizzey2021tunable}
Pizzey D 2021 {\em Rev. Sci. Instrum.\/} {\bf 92} 123002

\bibitem{alqarni2023device}
Alqarni S~A, Briscoe J~D, Higgins C~R, Logue F~D, Pizzey D, Robertson-Brown T~G and Hughes I~G 2023 {\em arXiv preprint arXiv:2308.16643\/}

\bibitem{trenec2011permanent}
Tr{\'e}nec G, Volondat W, Cugat O and Vigu{\'e} J 2011 {\em Appl. Opt.\/} {\bf 50} 4788--4797

\bibitem{sutter2020recording}
Sutter J~U {\em et~al.\/} 2020 {\em Comput. Electron. Agric.\/} {\bf 177} 105651

\bibitem{staerkind2023precision}
St{\ae}rkind H, Jensen K, M{\"u}ller J~H, Boer V~O, Petersen E~T and Polzik E~S 2023 {\em Phys. Rev. X\/} {\bf 13} 021036

\bibitem{staerkind2023high}
St{\ae}rkind H, Jensen K, M{\"u}ller J~H, Boer V~O, Polzik E~S and Petersen E~T 2023 {\em arXiv preprint arXiv:2309.12006\/}

\bibitem{higgins2021electromagnetically}
Higgins C~R and Hughes I~G 2021 {\em J. Phys. B: At. Mol. Opt. Phys.\/} {\bf 54} 165403

\bibitem{olsen2011optical}
Olsen B~A, Patton B, Jau Y~Y and Happer W 2011 {\em Phy. Rev. A\/} {\bf 84} 063410

\bibitem{ciampini2017optical}
Ciampini D, Battesti R, Rizzo C and Arimondo E 2017 {\em Phy. Rev. A\/} {\bf 96} 052504

\bibitem{auzinsh2022wide}
Auzinsh M, Sargsyan A, Tonoyan A, Leroy C, Momier R, Sarkisyan D and Papoyan A 2022 {\em Appl. Opt.\/} {\bf 61} 5749--5754

\bibitem{sargsyan2012hyperfine}
Sargsyan A, Hakhumyan G, Leroy C, Pashayan-Leroy Y, Papoyan A and Sarkisyan D 2012 {\em Opt. Lett.\/} {\bf 37} 1379--1381

\bibitem{weller2012optical}
Weller L, Kleinbach K~S, Zentile M~A, Knappe S, Hughes I~G and Adams C~S 2012 {\em Opt. Lett.\/} {\bf 37} 3405--3407

\bibitem{aplet1964faraday}
Aplet L~J and Carson J~W 1964 {\em Appl. Opt.\/} {\bf 3} 544--545

\bibitem{dick1991ultrahigh}
Dick D~J and Shay T~M 1991 {\em Opt. Lett.\/} {\bf 16} 867--869

\bibitem{yeh1982dispersive}
Yeh P 1982 {\em Appl. Opt.\/} {\bf 21} 2069--2075

\bibitem{gerhardt2018anomalous}
Gerhardt I 2018 {\em Opt. Lett.\/} {\bf 43} 5295--5298

\bibitem{kiefer2014faraday}
Kiefer W, L{\"o}w R, Wrachtrup J and Gerhardt I 2014 {\em Sci. Rep.\/} {\bf 4} 6552

\bibitem{logue2022better}
Logue F~D, Briscoe J~D, Pizzey D, Wrathmall S~A and Hughes I~G 2022 {\em Opt. Lett.\/} {\bf 47} 2975--2978

\bibitem{uhland2023build}
Uhland D, Dillmann H, Wang Y and Gerhardt I 2023 {\em New J. Phys.\/} {\bf 25} 125001

\bibitem{yin2022using}
Yin L, Qian D, Geng Z, Zhan H and Wu G 2022 {\em Opt. Express\/} {\bf 30} 36297--36306

\bibitem{faraday1846experimental}
Faraday M 1846 {\em Phil. Trans. R. Soc.\/}  1--20

\bibitem{zentile2014hyperfine}
Zentile M~A, Andrews R, Weller L, Knappe S, Adams C~S and Hughes I~G 2014 {\em J. Phys. B: At. Mol. Opt. Phys.\/} {\bf 47} 075005

\bibitem{chang2017faraday}
Chang P, Peng H, Zhang S, Chen Z, Luo B, Chen J and Guo H 2017 {\em Sci. Rep.\/} {\bf 7} 8995

\bibitem{voigt1899theorie}
Voigt W 1899 {\em Annalen der Physik\/} {\bf 303} 345--365

\bibitem{briscoe2023voigt}
Briscoe J~D, Logue F~D, Pizzey D, Wrathmall S~A and Hughes I~G 2023 {\em J. Phys. B: At. Mol. Opt. Phys.\/} {\bf 56} 105403

\bibitem{kudenov2020dual}
Kudenov M~W, Pantalone B and Yang R 2020 {\em Appl. Opt.\/} {\bf 59} 5282--5289

\bibitem{liu2023atomic}
Liu Z, Guan X, Qin X, Wang Z, Shi H, Zhang J, Miao J, Shi T, Dang A and Chen J 2023 {\em Appl. Phys. Lett.\/} {\bf 123}

\bibitem{muroo1994resonant}
Muroo K, Matsunobe T, Shishido Y, Tukubo Y and Yamamoto M 1994 {\em J. Opt. Soc. Am. B\/} {\bf 11} 409--414

\bibitem{mottola2023electromagnetically}
Mottola R, Buser G and Treutlein P 2023 {\em Phys. Rev. A\/} {\bf 108} 062820

\bibitem{mottola2023quantum}
Mottola R, Buser G and Treutlein P 2023 {\em Phys. Rev. Lett.\/} {\bf 131} 260801

\bibitem{budker2002resonant}
Budker D, Gawlik W, Kimball D~F, Rochester S~M, Yashchuk V~V and Weis A 2002 {\em Rev. Mod. Phys.\/} {\bf 74} 1153

\bibitem{auzinsh2010optically}
Auzinsh M, Budker D and Rochester S 2010 {\em {Optically polarized atoms: understanding light-atom interactions}\/} (Oxford University Press)

\bibitem{budker2000sensitive}
Budker D, Kimball D~F, Rochester S~M, Yashchuk V~V and Zolotorev M 2000 {\em Phys. Rev. A\/} {\bf 62} 043403

\bibitem{carr2020measuring}
Carr D~L, Spong N~L~R, Hughes I~G and Adams C~S 2020 {\em Eur. J. Phys.\/} {\bf 41} 025301

\bibitem{maxwell2021white}
Maxwell J~L, Hughes I~G and Adams C~S 2021 {\em Eur. J. Phys.\/} {\bf 43} 015302

\bibitem{wu1986optical}
Wu Z, Kitano M, Happer W, Hou M and Daniels J 1986 {\em Appl. Opt.\/} {\bf 25} 4483--4492

\bibitem{wolfenden1990use}
Wolfenden T~D, Baird P~E~G, Deeny J~A and Irie M 1990 {\em Meas. Sci. Technol.\/} {\bf 1} 1060

\bibitem{edwards1995precise}
Edwards N~H, Phipp S~J, Baird P~E~G and Nakayama S 1995 {\em Phys. Rev. Lett.\/} {\bf 74} 2654

\bibitem{edwards1995magneto}
Edwards N~H, Phipp S~J and Baird P~E~G 1995 {\em J. Phys. B: At. Mol. Opt. Phys.\/} {\bf 28} 4041

\bibitem{weller2012measuring}
Weller L, Dalton T, Siddons P, Adams C~S and Hughes I~G 2012 {\em J. Phys. B: At. Mol. Opt. Phys.\/} {\bf 45} 055001

\bibitem{kemp2011analytical}
Kemp S~L, Hughes I~G and Cornish S~L 2011 {\em J. Phys. B: At. Mol. Opt. Phys.\/} {\bf 44} 235004

\bibitem{siddons2010optical}
Siddons P, Adams C~S and Hughes I~G 2010 {\em Phys. Rev. A\/} {\bf 81} 043838

\bibitem{siddons2009gigahertz}
Siddons P, Bell N~C, Cai Y, Adams C~S and Hughes I~G 2009 {\em Nat. Photonics\/} {\bf 3} 225--229

\bibitem{rotondaro2015generalized}
Rotondaro M~D, Zhdanov B~V and Knize R~J 2015 {\em J. Opt. Soc. Am. B\/} {\bf 32} 2507--2513

\bibitem{palik1970infrared}
Palik E~D and Furdyna J~K 1970 {\em Rep. Prog. Phys.\/} {\bf 33} 1193

\bibitem{keaveney2018elecsus}
Keaveney J, Adams C~S and Hughes I~G 2018 {\em Comput. Phys. Commun.\/} {\bf 224} 311--324

\bibitem{titchmarsh1948introduction}
Titchmarsh E~C 1948 {\em {Introduction to the theory of Fourier integrals}\/} (Oxford University Press)

\bibitem{kanwal2013linear}
Kanwal R~P 2013 {\em {Linear integral equations}\/} (Springer Science \& Business Media)

\bibitem{zentile2015elecsus}
Zentile M~A, Keaveney J, Weller L, Whiting D~J, Adams C~S and Hughes I~G 2015 {\em Comput. Phys. Commun.\/} {\bf 189} 162--174

\bibitem{foot2004atomic}
Foot C~J 2004 {\em {Atomic physics}\/} vol~7 (Oxford University Press)

\bibitem{wellerThesis}
Weller L 2013 {\em PhD Thesis, Durham University\/} \urlprefix\url{http://etheses.dur.ac.uk/7747/}

\bibitem{siddons2009off}
Siddons P, Adams C~S and Hughes I~G 2009 {\em J. Phys. B: At. Mol. Opt. Phys.\/} {\bf 42} 175004

\bibitem{logue2023exploiting}
Logue F~D, Briscoe J~D, Pizzey D, Wrathmall S~A and Hughes I~G 2023 {\em arXiv preprint arXiv:2303.00081\/}

\bibitem{sherlock2009weak}
Sherlock B~E and Hughes I~G 2009 {\em Am. J. Phys.\/} {\bf 77} 111--115

\bibitem{keaveney2018automated}
Keaveney J 2018 {\em Rev. Sci. Instrum.\/} {\bf 89} 035114

\bibitem{alcock1984vapour}
Alcock C~B, Itkin V~P and Horrigan M~K 1984 {\em Can. Metall. Q.\/} {\bf 23} 309--313

\bibitem{hughes2010measurements}
Hughes I~G and Hase T 2010 {\em {Measurements and their uncertainties: a practical guide to modern error analysis}\/} (OUP Oxford)

\bibitem{data}
Briscoe J~D 2024 {Indirect measurement of atomic magneto-optical rotation via Hilbert transform [dataset]. Durham University Collections.} \url{http://doi.org/10.15128/r2x920fw90c}

\end{thebibliography}

\end{document}